\documentclass[3p,times]{elsarticle}

\usepackage{ecrc}

\volume{00}

\firstpage{1}

\journalname{Chaos, Solitons $\&$ Fractals}

\runauth{}

\jid{procs}

\jnltitlelogo{Chaos, Solitons $\&$ Fractals}

\CopyrightLine{2011}{Elsevier Ltd. All rights reserved}

\usepackage{amssymb}

\usepackage[figuresright]{rotating}
\usepackage{framed}
\usepackage{booktabs}

\begin{document}
®
\begin{frontmatter}




\dochead{}

\title{Anomalous Consistency in Mild Cognitive Impairment: A complex networks approach.}


\author[label1,label2]{J. H. Mart\'inez\corref{cor1}}
\cortext[cor1]{Corresponding author}
\ead{jh.martinez@alumnos.upm.es}
\ead[url]{http://johemart.wix.com/neurobelongings}

\author[label3]{P. Ariza}
\author[label4,label5]{M. Zanin}
\author[label6]{D. Papo}
\author[label7]{F. Maest\'u}
\author[label1]{Juan. M. Pastor}
\author[label7]{R. Bajo}
\author[label8]{Stefano Boccaletti}
\author[label9]{J. M. Buld\'u}

\address[label1]{Complex Systems Group, Technical University of Madrid, Madrid, Spain}
\address[label2]{Modelling and Simulation Laboratory, Universidad del Rosario de Colombia, Bogot\'a, Colombia}
\address[label3]{Laboratory of Biological Networks, Centre for Biomedical Technology (UPM-URJC), Madrid, Spain}
\address[label4]{Faculdade de Ci\^{e}ncias e Tecnologia, Departamento de Engenharia Electrot\'ecnica, Universidade Nova de Lisboa, Lisboa, Portugal}
\address[label5]{The INNAXIS foundation and Research Institute, Madrid, Spain}
\address[label6]{Computational Systems Biology Group, Centre for Biomedical Technology (UPM), Madrid, Spain}
\address[label7]{Laboratory of Cognitive and Computational Neuroscience, Centre for Biomedical Technology (UPM-UCM), Madrid, Spain}
\address[label8]{CNR-Istituto dei Sistemi Complessi, Via Madonna del Piano, 10, 50019 Sesto Fiorentino, Italy}
\address[label9]{Complex Systems Group, Universidad Rey Juan Carlos, Madrid, Spain}

\begin{abstract}
Increased variability in performance has been associated with the emergence of several neurological and psychiatric pathologies. However, whether and how consistency of neuronal activity may also be indicative of an underlying pathology is still poorly understood. Here we propose a novel method for evaluating consistency from non-invasive brain recordings. We evaluate the consistency of the cortical activity recorded with magnetoencephalography in a group of subjects diagnosed with Mild Cognitive Impairment (MCI), a condition sometimes prodromal of dementia, during the execution of a memory task. We use metrics coming from nonlinear dynamics to evaluate the consistency of cortical regions. A representation known as {\em parenclitic networks} is constructed, where atypical features are endowed with a network structure, the topological properties of which can be studied at various scales. Pathological conditions correspond to strongly heterogeneous networks, whereas typical or normative conditions are characterized by sparsely connected networks with homogeneous nodes. The analysis of this kind of networks allows identifying the extent to which consistency is affected in the MCI group and the focal points where MCI is especially severe. To the best of our knowledge, these results represent the first attempt at evaluating the consistency of brain functional activity using complex networks theory.

\end{abstract}


\end{frontmatter}


\section{Introduction}\label{intro}
Excessive variability in performance negatively impacts people's ability to carry out activities of daily living. Increased short-term fluctuations, particularly in reaction times, that cannot be attributed to systematic effects, such as learning, have been associated with a wide range of cognitive disorders including impaired top-down executive and attentional control processes, and with conditions including healthy ageing, and various neurological and psychiatric disorders ranging from Parkinson's disease \cite{camicioli2008,defrias2012}, multiple sclerosis \cite{bodling2012}, traumatic brain injury \cite{collins1996}, schizophrenia \cite{manoach2003} and various forms of dementia \cite{hultsch2000,tractenberg2011}. A number of studies attest an association between behavioral inconsistency and structural and functional brain abnormalities. For instance, diffusion tensor imaging showed a relationship between intra-individual variability in reaction times and white matter integrity, with variability increasing with white matter degradation, pathway connectivity degradation and brain dysfunction \cite{fjell2011,tamnes2012,teipel2010}. Behavioural inconsistency was also associated with neurotransmitter dysfunction, stress, and fatigue \cite{christensen2005,dixon2007,duchek2009,tractenberg2011,moy2011}.

However, whether and how behavioural consistency stems from a corresponding loss of consistency of functional brain activity is still unclear. For example, it has been shown that behavioural variability in the response time during a face recognition task negatively correlates with the brain signal variability, at least when children and adults are compared \cite{mcintosh2008}, the latter having lower response time variability combined with higher signal variability. In the current work we will focus on the variability of brain dynamics when carrying the same (memory) task. We study whether the dynamics of the recorded signal at diﬀerent cortical regions maintains or not its shape when the same memory task is carried out, and how the variability of each cortical region is related to that of other regions.

\begin{figure}[!hb]
\begin{center}
\includegraphics[width=0.7\textwidth]{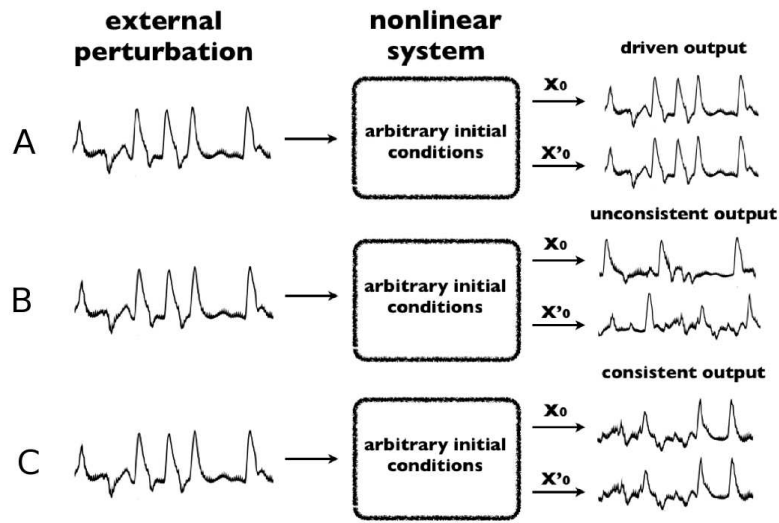}
\end{center}
\caption{
{\bf The phenomenon of consistency.} The consistency of a dynamical system relies on its ability to respond in the same way when the
the same external input is applied. In all panels (A, B, C), the same external perturbation is applied and the response
of the system with two different initial conditions ($\vec x_0$ and $\vec x'_0$) is (qualitatively) shown. 
In A, we show an example of a driven system, where the output always behaves in the same way (as the external perturbation) due to a strong forcing of the input. In B, the external perturbation is not
able to drive the output which responds in a different way when the same input is applied (due to different initial conditions).
In C, we show an example of consistent dynamics. When the same input is applied, the system's output always
behaves in the same way, despite not mimicking the driving dynamics.
}
\label{fig01}
\end{figure}

In physics, {\em consistency} \cite{uchida2004} has been studied with a series of different dynamical models \cite{goldobin2008,uchida2008,perez2011}. The emergence of a consistent response requires a high synchronization between different outputs of a nonlinear system (i.e., different realizations with different initial conditions) when the same external input is applied. Nevertheless, consistency does not imply the observation of a synchronized state between the external input and the system's response as it is the case of an entrained or driven system.  Figure \ref{fig01} shows, qualitatively, the difference between a driven system and a consistent/inconsistent system.

Here we propose a new methodology for quantifying neuronal consistency that can be applied to noninvasive diagnostic procedures. This method involves constructing a {\em parenclitic network} representation \citep{zanin2011,zanin2013} based on anomalous behaviour of a certain set of features. In our case, each node of the parenclitic network, each of them corresponding to a certain cortical region, is associated with a feature: its dynamical consistency during a memory task. We construct a network where nodes (i.e., cortical regions) have a link between them with a weight that depends on how their consistency diverges from an expected value. The topological characteristics of the parenclitic networks can be used to extract information about how consistency is lost/maintained across the whole functional network. In such a representation, atypical or pathological conditions lead to the existence of a high number of links with large weights. Previous applications of this methodology to biological data have shown that parenclitic networks obtained from datasets associated with pathological conditions lead to strongly heterogeneous networks, i.e., networks where some nodes have a degree that is much higher than expected for links created at random, whereas typical or normative conditions are characterized by sparsely connected networks with homogeneous nodes \citep{zanin2011,zanin2013}. In essence, a parenclitic representation equips the set of abnormalities of a system with a network characterization with topological properties at various scales.

Specifically, we study the consistency of functional brain activity in a group of patients diagnosed with mild cognitive impairment (MCI). MCI is a clinical condition in which subjects experience memory loss to a greater extent than  would be expected for age, who while not meeting the criteria for clinically probable Alzheimer's Disease (AD) are nonetheless at increased risk of developing it. Behavioural evidence shows that, compared to cognitively healthy ageing, MCI has been associated with increased response time variability and particularly in those subjects later developing AD \citep{christensen2005,dixon2007,gorus2008,duchek2009,fjell2011,tamnes2012}. Abnormal consistency in MCI may therefore represent a measure of functional integrity that may help identifying those patients who ultimately lapse into fully-ﬂedged dementia.

On the other hand, recent studies have shown how the functional networks of MCI patients obtained during a memory test show differences in their topologies when compared to healthy subjects \citep{SL-bajo,buldu2011}. This fact indicates that a global reorganization of brain activity is occurring, raising the question of how the consistency of a brain network is affected by this reorganization. With the aim of studying the consistency of the whole brain network, we used magnetoencephalography (MEG) to record the brain activity of a group of patients suffering from MCI and a healthy control group as they carried out the Sternberg short-term memory task. We then computed the consistency of each brain site for each individual of each group and finally constructed the parenclitic network for the differences in consistency between MCIs and controls (see Materials and Methods for details). The structure of the parenclitic networks highlights those regions whose consistency is most affected by the disease and suggests ways in which the effects of MCI may propagate through the functional brain network.

\section{Materials and Methods}\label{sec:1}
\subsection{Subjects}\label{subsec:11}
All subjects or legal representatives provided written consent to participate in the study, which was approved by the local ethics committee of the Hospital Cl\'inico Universitario San Carlos (Madrid, Spain). Fourteen right handed patients with MCI were recruited from the Geriatric Unit of the ``Hospital Cl\'inico Universitario San Carlos Madrid''. In addition, fourteen age-matched elderly participants without MCI were included as the control group. 
In addition to age, years of education were matched to the MCI group (for details see \citep{SL-bajo}): 10 years for the MCI group and 11 years for controls. To confirm the absence of memory complaints, a score of 0 was required in a 4-question questionnaire \citep{mitchell}. None of the participants had a history of neurological or psychiatric condition.
The diagnosis of MCI was made according to the criteria proposed by Petersen 
\citep{grundman,petersen}. None of the subjects with MCI had evidence of depression as measured using the geriatric depression scale (score lower than 9) \citep{yesavage}.
MCI subjects and healthy participants underwent a neuropsychological assessment in order to establish their cognitive status with respect to multiple cognitive functions. Specifically, memory impairment was assessed using the Logical Memory immediate (LM1) and delayed (LM2) subtests of the Wechsler Memory Scale-III-Revised. Two scales of cognitive and functional status were applied as well: the Spanish version of the Mini Mental State Exam (MMSE) \citep{lobo}, and the Global Deterioration Scale/Functional Assessment Staging (GDS/FAST). A summarized demographical information is stored in Tab. \ref{tab:01}.

\begin{table}[!hb]
\caption{Demographic and clinical information of the Control and MCI groups \citep{SL-bajo}. \textbf{MMSE}: Mini Mental State Exam (maximum score is 30); \textbf{GDS}: Global Deterioration Scale; \textbf{LM1}: Logical Memory immediate recall; \textbf{LM2}: Logical Memory delayed recall}
\label{tab:01}       
\renewcommand{\arraystretch}{1.5}
\centering
\begin{tabular}{llllllllllll}
\hline
\noalign{\smallskip}
\multicolumn{3}{l}{\textbf{GDS}} & \multicolumn{3}{l}{\textbf{LM1}} & \multicolumn{3}{l}{\textbf{LM2}} \\
\noalign{\smallskip}
\cline{1-2}
\cline{4-5}
\cline{7-8}
\noalign{\smallskip}
Controls & MCIs & & Controls & MCIs & & Controls & MCIs & \\
\noalign{\smallskip}
\bottomrule[1.2pt]
\noalign{\smallskip}
1 & 3 & & $42.5\pm8$ & $19.1\pm5$ & & $26.7\pm7$ & $13.1\pm6$ & \\
\noalign{\smallskip}
\hline
\end{tabular}\\
\begin{tabular}{llllllllllll}
\hline
\noalign{\smallskip}
\multicolumn{3}{l}{\textbf{Sample(Sex)}} & \multicolumn{3}{l}{\textbf{Age}} & \multicolumn{3}{l}{\textbf{MMSE}} \\
\noalign{\smallskip}
\cline{1-2}
\cline{4-5}
\cline{7-8}
\noalign{\smallskip}
Controls & MCIs & & Controls & MCIs & & Controls & MCIs &  \\
\noalign{\smallskip}
\bottomrule[1.2pt]
\noalign{\smallskip}
14(9 Female) & 14(9 Female) & & $70.6\pm8.1$ & $74.7\pm3.6$ & & $26.75\pm0.9$ & $25\pm1$ & \\
\noalign{\smallskip}
\hline
\end{tabular}
\end{table}

\subsection{Task}\label{subsec12}
Magnetoencephalography (MEG) scans were obtained in the context of a modified version of the Sternberg test \citep{sternberg1966}, which is a well-known task for studying working memory. Specifically, it consisted of a letter-probe test \citep{taskT-deToledo,taskT-maestu} in which a set of five letters was presented and subjects were asked to keep the letters in mind. After the presentation of the five-letter set, a series of single letters (1000 ms in duration) was introduced one at a time, and participants were asked to press a button with their right hand when a letter of the previous set was detected. 
The list consisted of 250 letters in which half were targets (previously presented letters), and half distracters (not previously presented letters). Subjects undertook a training series before the actual test and the task did not start until the participant demonstrated that he/she could remember the five letter set.  
Subjects' responses were classified into four different categories: hits, false alarms, correct rejections and omissions. Only hits were considered for further analysis because the authors were interested in evaluating the functional connectivity patterns which support recognition success. The percentage of hits (80\% control group and 84\% MCI group) and correct rejections (92\% control group and 89\% MCI group) was high enough in both groups, indicating that participants actively engaged in the task. No significant difference between the two groups during the memory task was revealed. Note that despite patients suffering from MCI have deficits in episodic memory, previous results have shown differences in the functional networks obtained during a short-term memory tests \citep{SL-bajo,buldu2011} and even at resting state \citep{seo2013}. Therefore, due to the unpredictability of the episodic memory lags, we have studied the effects of the disease during a short-term memory test.

\subsection{MEG recordings}\label{subsec13}
MEG signals were recorded with a 254 Hz sampling frequency and a band pass of 0.5 to 50 Hz, using a 147-channel whole-head magnetometer (MAGNES 2500 WH, 4-D Neuroimaging) confined in a magnetically shielded room. An environmental noise reduction algorithm using reference channels at a distance from the MEG sensors was applied to the data. 
Letters of the Sternberg test were projected through a LCD videoprojector (SONY VPLX600E), situated outside of a magnetically-shielded room, onto a series of in-room mirrors, the last of which was suspended approximately 1 meter above the participant's face. The letters subtended 1.8 and 3 degrees of horizontal and vertical visual angle respectively.
Prior to functional connectivity analysis, all recordings were visually inspected by an experienced investigator, and all of them containing visible blinks, eye movements or muscular artifacts were excluded from further analysis. Thirty-five epochs (1 second each one) were finally used. This lower bound was determined by the participant with least epochs, in order to have an equal number of epochs across participants. This way, we randomly selected thirty-five epochs from those individuals with a higher number of valid epochs.

\subsection{Analysis}\label{subsec14}
The first step is to quantify how consistent the output of each channel is. This can be done by computing how coherent the outputs of the same
channel are when the same task is carried out. With this aim, for each individual, we calculate the Synchronization Likelihood (SL) \citep{SL-stam,SL-bajo}  between each pair of MEG time series within the same channel, considering only successful identifications of the target letters. SL is a nonlinear measure of correlation indicating synchronization for values close to one and the absence of any correlation when approaching zero. 
While other measures to evaluate linear or nonlinear correlation between time series could have been used \citep{pereda2005}, SL has proven an adequate measure for capturing the interdependencies between MEG time series obtained during a short-term memory tests \citep{buldu2011}.

This way, we evaluate if the cortical activity measured at each channel during a positive identification of a letter is consistent, e.g., has similar temporal evolution when repeating the same task, despite the initial conditions are intrinsically different. We calculate the average of the pairwise SL of all non-repeated permutations of the 35 time series recorded within each magnetometer in order to get the Channel Consistency (CC). Note that the 35 time series of each channel are not combined but compared between them to extract the CC.

At this stage, we have a dataset based on 147 CC values for the 14 subjects of the two different groups (MCIs and controls), which were used to build the corresponding parenclitic networks \citep{zanin2013}.
\begin{figure}[!h]
\begin{center}
\includegraphics[width=1\textwidth]{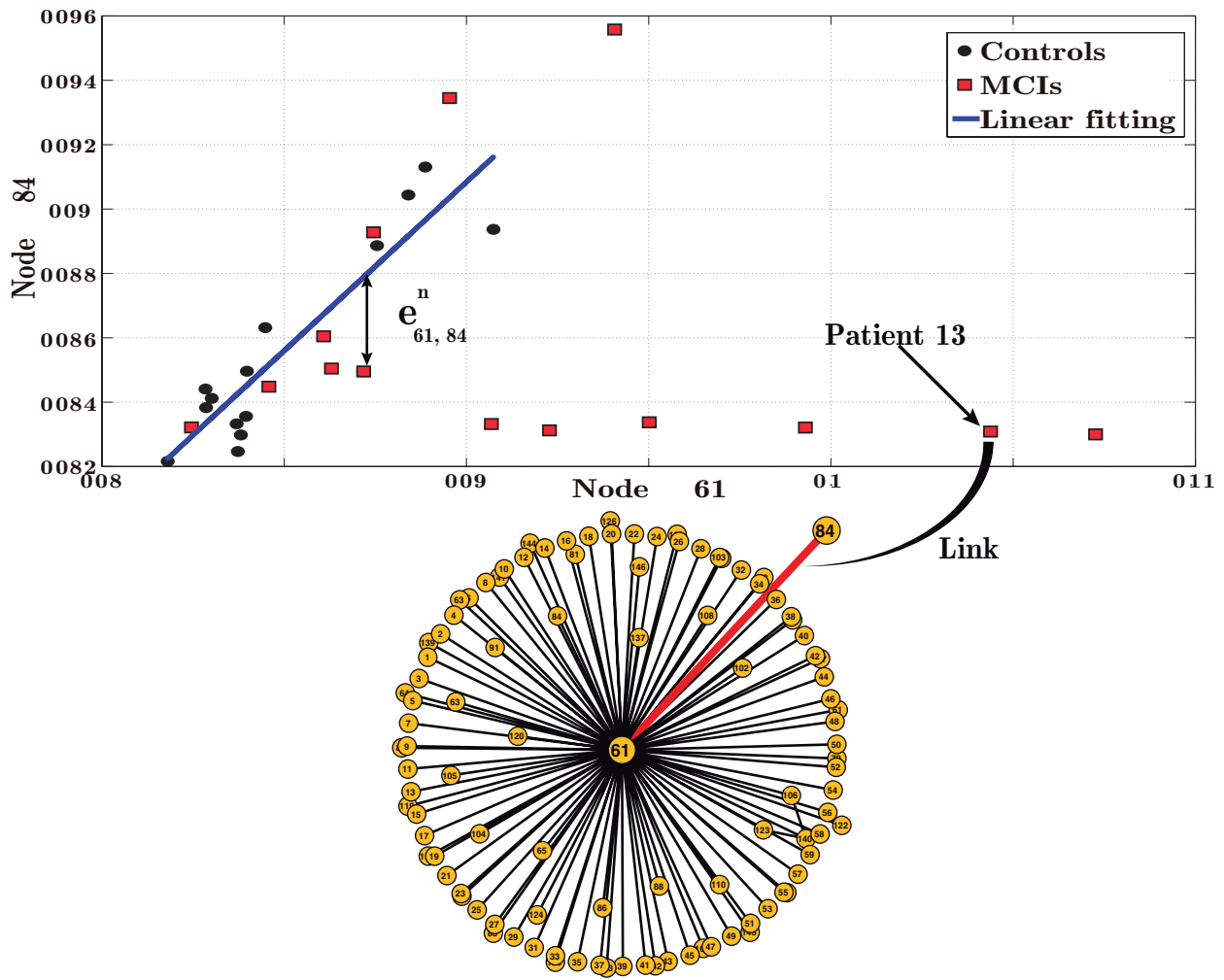}
\end{center}
\caption{
{\bf Creating a parenclitic network.} The {\em channel consistency} (CC) of channels 61 and 84 (each one corresponding to a node of the parenclitic network) is plotted for the 14 controls (black circles) and the 14 patients suffering from MCI (red squares). A linear fitting of the CC pairs of the control group
is calculated (blue line) and it will be taken as the reference for a normal behaviour. $e^{n}_{61,84}$ accounts for the deviation of CC from the reference value for the individual $n$. Each individual, will have its own parenclitic networks, where the weight of the link between nodes $61$ and $84$ is calculated as $Z^n_{61,84}=|e^n_{61,84}|/ \sigma_{61,84}$, being $\sigma_{61,84}$ the
standard deviation of the CC values from the reference line. This way, the larger the weight of the link the higher the deviation from the reference given by the control group. This procedure is repeated for each pair of channels, leading to a parenclitic network for each individual, whose links quantify how far is the consistency of a pair of channels from the normal behaviour. The bottom plot shows part of the parenclitic network of patient $13$, specifically the neighborhood of node $61$ once all pairs of channels underwent the same procedure.
}
\label{fig02}
\end{figure}

The method for the network construction is explained in Fig. \ref{fig02}, which shows a specific example with channels $i=61$ and $j=84$. Black dots represent the CC of these two magnetometers for all control subjects and the blue line is their corresponding linear fitting. Note that there is no previous evidence that a linear correlation between the consistency of two independent channels exists, since, to the best of our knowledge, this is the first study obtaining networks from intra-channel consistency. Nevertheless, as we will see, assuming a linear correlation leads to a clear distinction between groups, based on statistical significant differences between the network parameters of the control and MCI groups. Interestingly, assuming a correlation described by a second-order polynomial leads to the same qualitative results (not shown here). This is due to the fact that the number of points used to estimate the correlation function (fourteen in the absence of outliers) is low, since it is difficult to have large datasets of individuals. A combination of both kinds of correlations together with other non-linear methods to evaluate functional dependencies between nodes deserves its own study and it will probably depend on the kind of data and problem (disease) under investigation.

The errors of the control group are adjusted to a normal distribution in order to obtain its standard deviation $\sigma_{i,j}$. Once the standard deviation is obtained, we recalculate the correlation function excluding the outlier points of the control group. Individuals of the control group with an error higher than $2\sigma_{i,j}$ are assumed to be outliers and are not considered for the definition of the correlation function. This way we exclude 6030 outliers from all the $N(N-1)/2$ correlation diagrams, which represent the 4.2\% of the total number of points. Importantly though, the outliers are considered when calculating the average properties of networks of the control. This allows taking into account variability inside the control group.

Red squares correspond to the CC of the same
pair of channels for the 14 MCI patients. The deviation from the expected value given by the linear fitting is 
designated as the error $e_{i,j}$ of the joint consistency of channels $i$ and $j$. The z-score of each pair of channels associated 
to a subject $n$ will measure how far the consistency of both channels is from the expected value and it is obtained as 
$Z^n_{i,j}=|e^n_{i,j}|/ \sigma_{i,j}$. 
The next step is to project the z-score dataset into a parenclitic network. 
In general, the links of a parenclitic network are created with a weight that is proportional to the deviations of a 
certain feature from an expected value \citep{zanin2013}. These networks are weighted and non-directed, and they unveil 
important topological differences between a reference group and a group with a certain anomaly \citep{zanin2011}. In our case, the nodes 
of the network will be the channels measuring the activity of a certain cortical region and the links between a pair of 
nodes $i$ and $j$ will be the z-score $Z^n_{i,j}$ measuring their deviation from its expected value. 

The procedure followed to obtain the links and weights of  nodes $i=61$ and $j=84$ is repeated for the CC of the $147\times146/2$ possible pairs of channels in order to obtain all links of the networks. This way, we obtain a weighted matrix which can be represented as a fully connected network.
Finally, we need to apply a threshold in order to consider only those deviations that are relevant enough. The thresholding process consists in 
considering only the $L$ links with higher weight, leading to a sparse
matrix whose topology will be further analyzed. Nevertheless, it is a delicate step, since a very low threshold will maintain spurious data that may hide the observation of the real network structure, while a very high threshold could dismiss valuable information.
In order to adequately set the threshold value, we repeated the analysis for different values of $L$ and calculated the corresponding network parameters. Next, we identify the threshold that showed more differences between the network parameters of the control and the MCI groups \citep{zanin2012}. Specifically, we focused on the differences in the local $\bar E_l$ and global $\bar E_g$ efficiency, since the dependence of these parameters on the deletion of links is smoother than the clustering coefficient C of the shortest path length d \citep{latora-Eloc}, and, in turn, showed a maximum difference around the same value of L.

After following this procedure, we finally consider only the $L=400$ links with the largest  $Z^n_{i,j}$ of each parenclitic network. This way, we obtain a set of weighted sparse networks for the control and MCI subjects with the same number of nodes $N$ and links $L$. By computing a set of network metrics \citep{newman2010},  we are able to compare whether the networks differ in their topological organization and what are the kind of network structures associated to each group. All network metrics were statistically analyzed on the basis of the mean differences between both populations and 5000 different permutations were performed in order to obtain the corresponding $p$ value of each network metric.

\section{Results}\label{sec2}		
We characterize the following properties of the parenclitic networks: a) the degree and strength
of the nodes, b) the clustering coefficient, c) the characteristic path length, d) the local and global efficiency and e) the eigenvector centrality of the nodes (i.e., the identification of the network hubs).
\subsection{Degree, strength and hubs}\label{subsec21}
Our first inspection of the network topology focuses on the local properties of its nodes. Specifically, we computed the highest degree $K_{max}$ of the network, which gives the highest number of connections a node has. $K_{max}$ is an indicator of the existence of network hubs, i.e., nodes with a number of links much higher than the expected from the average connectivity.

If we take into account the weight associated to the links, we can also compute the node strength $S(i)$, which is given by the sum of all weights of the links attached to node $i$.
The maximum strength $S_{max}$ and the average strength $\bar{S}$ of the networks are obtained as the maximum/average of $S(i)$ over all nodes. Note that Smax better captures the importance of a node in the network, since apart from depending on the number of links a node has, it takes into account the weights associated with these links.

\begin{figure}[!hb]
\begin{center}
\includegraphics[width=0.6\textwidth]{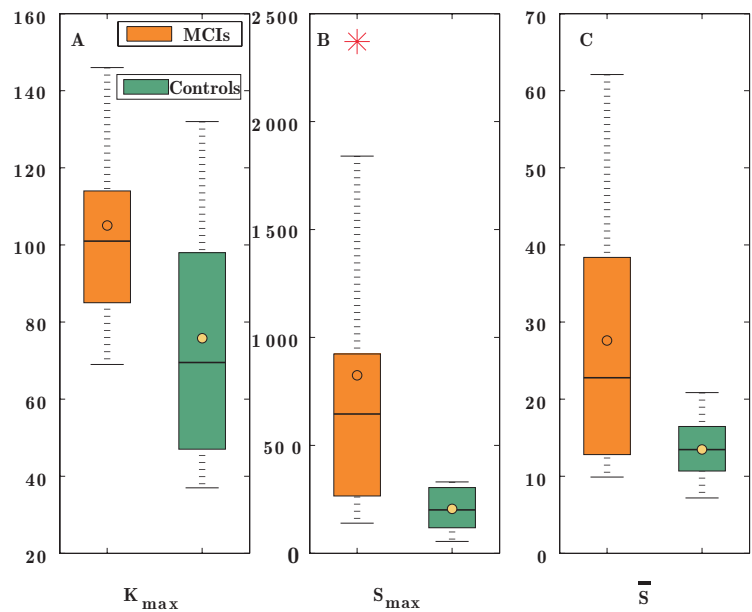}
\end{center}
\caption{
{\bf Highest degree, maximum strength and average strength.}
From left to right, box \& whisker representation showing the first, second, third quartile and the average of: 
(A) the highest degree $K_{max}$, (B) maximum strength $S_{max}$ and (C) average strength $\bar S$.
Patient and control groups are orange and green, respectively. Red stars account for outlier values.
}
\label{fig03}
\end{figure}

 Figure \ref{fig03} shows a comparison of the highest degree $K_{max}$, maximum strength $S_{max}$ and the average 
strength $\bar{S}$ of the control and MCI groups. We use the {\em box} \& {\em whisker} representation which highlights the main 
statistical quantities of the datasets, i.e., the first, second and third quartile and the mean. When looking at the highest degree $K_{max}$ we can see how, in the MCI group, the mean, median (or second quartile), third quartile and, in general, all values are around 50 \% higher than the corresponding values of the control group (Fig. \ref{fig03}A). This finding evidences the presence of hubs with higher number of links in the
parenclitic network associated to the MCI group. Since the number of links is limited to $L=400$ in both groups, the fact that large hubs arise in the MCI networks also reveal the formation of more heterogeneous structures.

Figure \ref{fig03}B shows the maximum strength $S_{max}$ for each group. 
Red stars are outliers that show that the maximum strengths of the MCI group follow not a normal but a skewed distribution. The behaviour of $S_{max}$ remains similar to $K_{max}$ and evidences that the existence of stronger hubs 
in the MCI group is reinforced when considering the weight of the links. Therefore both measures indicate the existence of certain nodes that accumulate large deviations in their expected value of consistency, leading to more heterogeneous networks.

In Fig. \ref{fig03}C we plot the average strength $\bar{S}$ of the networks to confirm that the MCI patients have higher
deviations from the reference value than the control individuals. As expected, the average strength of the MCI group $\bar{S}^{MCI}=27.59$ is much higher than the control group ($\bar{S}^{control}=13.49$), since the node strength accumulates the errors of all its links
(see Materials and Methods). The higher the value of $\bar{S}$, the larger the deviation of the overall consistency of the functional network. It is important to remark the difference between the maximum and average strength of both groups, which is 
much higher in the MCI group. This fact reveals the more heterogeneous structure of the MCI networks, where nodes with strong deviations
from the expected value arise (i.e. the higher the strength of a node, the more anomalous its consistency is).
\begin{figure}[!hb]
\begin{center}
\includegraphics[width=0.75\textwidth]{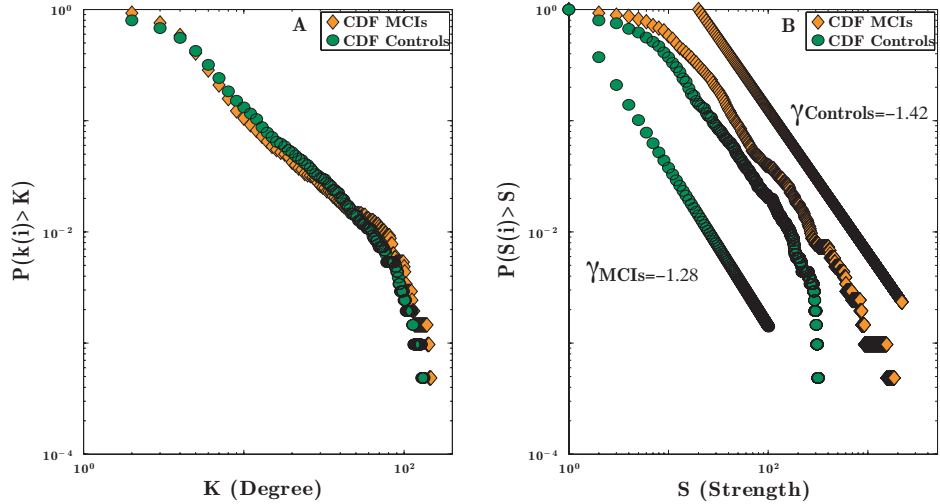}
\end{center}
\caption{
{\bf CDF of the degree and strength distributions.} (A) Cumulative Distribution Functions (CDF) of the probability of finding a node
with a degree (strength in (B)) higher than $k$ ($S$ in (B)). Green circles correspond to the control group and orange circles to the MCI group.
When the strength of the nodes is considered (B), we obtain the power law distribution $P(S(i)\geq S) \sim S^{-\gamma}$ in both cases, as indicated
by a straight line in the log-log scale.
}
\label{fig04}
\end{figure}

Finally, we calculate the cumulative distribution function (CDF) of the degree and strength of the nodes both in the control and MCI groups. For each node $i$ (i.e., cortical region), we compute its corresponding average degree $\langle k(i) \rangle$. Next, we obtain the CDF by computing the percentage of nodes with a degree $\langle k (i)\rangle \geq k$. The same methodology 
is followed to calculate the CDF of the node strength $P(\langle S(i)\rangle \geq S)$. Both distributions are plotted in Fig. \ref{fig04}  with regard to the subjects of each group. 
Permutation t-test for the degree and strength CDFs was performed. Taking into account 5000 randomizations we found significant statistically differences in the strength distributions with a  p-value = 0.004. Differences in the degree CDFs were not statistically significant (p-value = 0.79).

We can observe how the degree distributions cross at $k_c=40$ showing that nodes with a degree higher than $k_c$ are
more probable in the MCI group than in the control group. This confirms what the values of Kmax and Smax already suggested: a number of nodes in the parenclitic networks of the MCI group grossly deviate from their expected consistency.

The strength distribution spreads over three orders of magnitude and allows the CDF to show a power law decay $\sim k^{-\gamma}$. The exponent of the power law is lower in the MCI group ($\gamma^{MCI}=1.28<\gamma^{control}=1.42$) which reveals the existence of hubs with a larger strength. These hubs play a relevant role in the structure of the network, since they accumulate a higher percentage of the link weights: they are the core of the divergences with respect to the normal (healthy) values of consistency. 

\subsection{Local vs. global properties}\label{subsec22}
\begin{figure}[!hb]
\begin{center}
\includegraphics[width=0.6\textwidth]{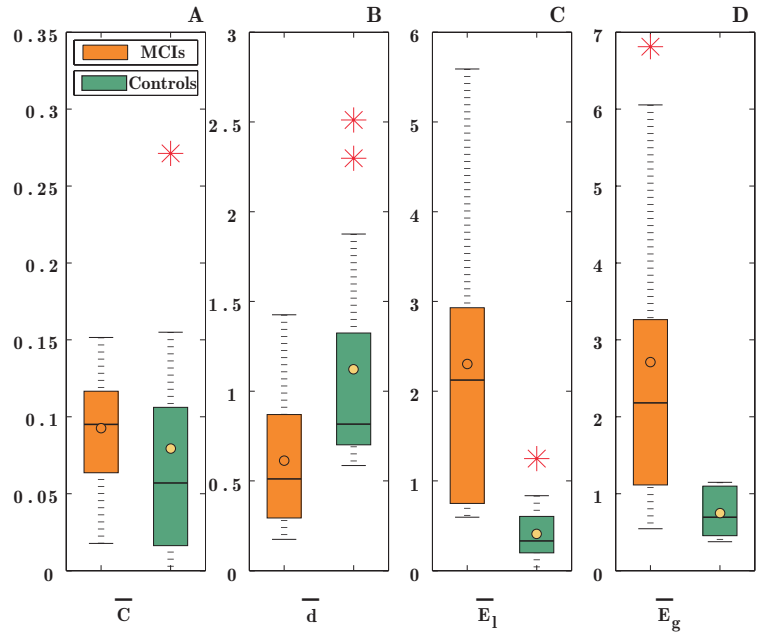}
\end{center}
\caption{
{\bf Clustering, shortest path length, local efficiency and global efficiency.}
Box \& whisker representation of: (A) the clustering $\bar C$ , (B) shortest path length $\bar d$, 
(C) local efficiency $\bar E_l$ and (D) global efficiency $\bar E_g$.
Orange and green bars correspond, respectively, to the MCI and control groups. Red stars are the outlier values. P-values of the network parameters are given in Table \ref{tab:02}.
}
\label{fig05}
\end{figure}

Next, we calculate two local (clustering coefficient $\bar C$ and local efficiency $\bar E_l$) and two global (average path length $\bar L$ and global efficiency $\bar E_g$) properties of the network.
The clustering coefficient $c_i$ of a node $i$ is the number of triangles around a node (i.e., number of neighbors that, 
in turn, are neighbors between them) divided by the highest possible number of triangles 
(i.e., the number of triangles if all of its neighbors were connected between them) \citep{watsStrogatz}. $c_i$ is calculated using a generalization of this metric for weighted networks \citep{onnelaWeighted}, and averaged over the whole network to obtain the clustering coefficient $\bar C$ per individual. $\bar C$ is an indicator of the local density of connections inside the network, and it has been related to the local resilience of a network against removal of links (i.e., the highest the clustering, the highest the local resilience) \citep{bocaletti2006}. Figure \ref{fig05}A shows $\bar C$ for the two groups under study. We can observe how the MCI network has a largest clustering coefficient, indicating a higher density of connections at the local level. Interestingly, the clustering coefficient is also an indicator of the network randomness since random networks have a value of $\bar C$ close to zero. Thus, the lowest value of $\bar C$ of the control group indicates that its network topology is closer to a random structure.

Now let us have a look at a global property of the network: the average shortest path $\bar d$, i.e., the average number of steps to go from a node to any other.
To obtain the value of $\bar d$ we first calculate the distance matrix $D$ for all parenclitic networks. We assign a length to each
link as $l_{i,j}=1/{Z_{i,j}}$, i.e. the higher the weight of the link, the shorter the ``distance" between node $i$ and $j$. 
Each $d_{i,j}$ element of the $D$ matrix is the shortest path between nodes $i$ and $j$ (i.e., the lowest combination of links' lengths to go from $i$ to $j$), which is calculated 
using the Dijkstra's algorithm \citep{dijkstraAlgor}. Finally, the average shortest path $\bar d$ is just the average of all elements of matrix $D$. 
Figure \ref{fig05}B shows that the MCI group has a lower value of $\bar d$, which is a consequence of having higher weights (i.e., shorter distances) in its connectivity matrix (since, as we have seen, $\bar S^{MCI}>\bar{S}^{control}$). Parenclitic networks are capturing how alterations of the expected consistency are distributed over the whole network and, therefore, the low value of $\bar d$ reveals that the loss of consistency propagates with a shorter number of steps in the MCI group.
That is not good news for the resilience of the consistency when MCI emerges.\\

Both parameters $\bar C$ and $\bar d$ can be reinterpreted in terms of how efficient is the network when transmitting information from 
one node to any other in the network. With this aim, Latora et al. \citep{latora-Eloc} introduced the concept of local $\bar E_l$ and global
$\bar E_g$ efficiency, which accounts for the harmonic mean of the inverse of the number of steps between any pair of nodes. 
Efficiency can be defined at a local level, i.e., within the community of neighbours of a certain node, 
or globally, i.e. over the whole network. This way, high values of local/global efficiency indicate a good transmission of information 
(at the local/global scale) in terms of the number of steps.
Figures \ref{fig05}C-D show the local $\bar E_l$ and global $\bar E_g$ efficiency for both groups. We can observe how in both cases
the efficiency is higher in the MCI networks, which indicates that the network of dysfunctions is better organized. 

Table \ref{tab:02} summarizes the average of all network metrics. Note that despite the classical definition of $C$, $d$, $E_l$ and $E_g$ constrain the values of these to the interval [0,1] \citep{newman2010}, the fact that we are using weighted connectivity matrices, which contain more information about the interdependency between nodes, leads to values that can exceed this range \citep{rubinov2010}. A comparison between control and MCIs group, for each network parameter, was developed via a non parametric Kruskal Wallis test, where we have computed the $p$-values (5000 permutations each) that illustrate how 
significant the statistical differences are.

\begin{table}[!hb]
\caption{Summary of the network metrics for the control and MCI groups: Highest degree $\mathbf{K_{max}}$,  maximum strength $\mathbf{S_{max}}$, average strength $\mathbf{\bar{S}}$, clustering $\mathbf{\bar{C}}$, average shortest path $\mathbf{\bar{d}}$, local efficiency $\mathbf{\bar{E}_l}$ and global efficiency $\mathbf{\bar{E}_g}$. The p-value of each metric is also indicated.}
\label{tab:02}       
\renewcommand{\arraystretch}{1.5}
\centering
\begin{tabular}{lllllllll}
\hline
\noalign{\smallskip}
\multicolumn{3}{l}{$\mathbf{K_{max}}$ $(p=0.016)$} & \multicolumn{3}{l}{$\mathbf{S_{max}}$ $(p=0.0002)$} & \multicolumn{3}{l}{$\mathbf{\bar S}$ $(p=0.004)$} \\
\noalign{\smallskip}
\cline{1-2}
\cline{4-5}
\cline{7-8}
\noalign{\smallskip}
Controls & MCIs &  & Controls & MCIs &  & Controls & MCIs &  \\
\noalign{\smallskip}
\bottomrule[1.3pt]
\noalign{\smallskip}
75.78 & 105 & & 206.27 & 824.30 & & 13.49 & 27.59 &  \\
\noalign{\smallskip}
\hline
\end{tabular}
\begin{tabular}{llllllllllll}
\hline
\noalign{\smallskip}
\multicolumn{3}{l}{$\mathbf{\bar C}$ $(p=0.562)$} & \multicolumn{3}{l}{$\mathbf{\bar d}$ $(p=0.012)$} & \multicolumn{3}{l}{$\mathbf{\bar E_l}$ $(p=0.0002)$} & \multicolumn{3}{l}{$\mathbf{\bar E_g}$ $(p=0.0002)$} \\
\noalign{\smallskip}
\cline{1-2}
\cline{4-5}
\cline{7-8}
\cline{10-11}
\noalign{\smallskip}
Controls & MCIs & & Controls & MCIs & & Controls & MCIs & & Controls & MCIs & \\
\noalign{\smallskip}
\bottomrule[1.2pt]
\noalign{\smallskip}
 0.08 &  0.09 & &  1.12 &  0.61 & & 0.40 & 2.30& & 0.74 &  2.70 & \\
\noalign{\smallskip}
\hline
\end{tabular}
\end{table}

\subsection{Localizing focal nodes}\label{subsec23}
\begin{figure}[!hb]
\begin{center}
\includegraphics[width=1\textwidth]{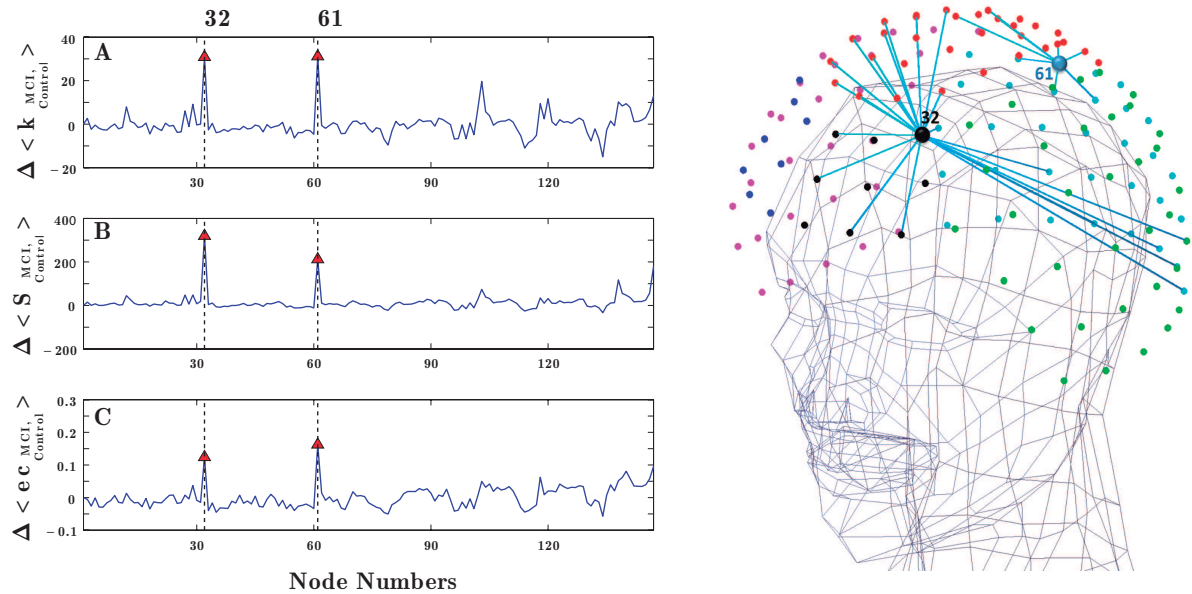}
\end{center}
\caption{
{\bf Localizing focal nodes in the consistency impairment.} We calculate the differences in the node degree $k(i)$ (A), node strength $s(i)$ (B) and eigenvector centrality $ec(i)$ (C). In all cases, nodes 32 and 61 accumulate the highest variations, indicating that are the nodes whose consistency is affected the most by the disease. On the right plot, we show the position and the local 
network of connections of these two nodes, where only the $30$ links with higher weights have been plot.
}
\label{fig06}
\end{figure}

Parenclitic networks allow detecting those nodes whose features (consistency in our case)
diverge the most from the expected behaviour. This task is carried out by finding the network hubs
and quantifying their importance.
With this aim, we calculate the degree $k(i)$, strength $S(i)$
and eigenvector centrality $ec(i)$ of nodes belonging to both populations, all these metrics commonly used as quantifiers
of the network hubs. As explained in the previous section,
the degree and strength account respectively for the number of links and total accumulated weight of a node. Both 
metrics rely on the local properties of the nodes, which is not the case of $ec(i)$. Eigenvector centrality is a global
measure of importance that takes into account not only the number of connections/weights of a node, but the number
of connections/weights of its neighbors \citep{newman2010}. $ec(i)$ is calculated from the eigenvector associated to the largest
eigenvalue of the weighted connection matrix $W$ whose elements are, in our case, $Z_{i,j}$.
We proceed as follows: two vectors $\vec k_{MCI}$ and $\vec k_{control}$ of length $N=147$ (one element per node) 
contain the average degree of the nodes of each specified group. 
The difference of the elements of both vectors $\Delta\langle \vec k_{MCI,control}\rangle=\vec{k}_{MCI}-\vec{k}_{control}$ accounts
for the difference of node degree between both groups and reflects what nodes increase (or decrease) their importance in the network. Figure \ref{fig06}A shows $\Delta\langle \vec k_{MCI,control}\rangle$, where two peaks stand out over the rest of the degree variations. Nodes 32 and 61
have a much higher degree in the MCI parenclitic networks than in the control ones. This fact indicates that these two
nodes accumulate the majority of variations related to the consistent behaviour.
In a similar way, we obtain the variations of the strength $\vec S_{MCI,control}$ and eigenvector centrality $\vec ec_{MCI,control}$.
Figure \ref{fig06}B-C shows the difference between groups of these two metrics $\Delta\langle \vec S_{MCI,control}\rangle$ (B),
$\Delta\langle \vec{ec}_{MCI,control}\rangle$ (C). Independent of the metric, again two peaks 
appear at nodes $32$ and $61$, confirming that they are the nodes whose consistency is affected the most by the disease.
It is worth noting that these two nodes are not necessarily those nodes whose consistency increased/decreased the most. Node 61 has an 8.14\% variation in its consistency (\#4 in the raking of consistency variations) and node 32 around 2.71\% (\#25 in the consistency variation ranking). This fact indicates that parenclitic networks go beyond the local changes of consistency and account for the way variations affect the interplay, based on consistency, between nodes.

In Fig. \ref{fig06} we plot a 3D representation of these {\em focal nodes} together with their local network of interactions.
This plot shows the local basin of influence of the network hubs and gives an idea about where the disease is being more severe,
at least when consistency is taken into account. It is valuable to compare the position of the most affected nodes with previous results on how MCI affects functional networks. In Buldú et al. \citep{buldu2011}, it was shown that both the frontal and the occipital lobes contain those nodes of the functional network whose synchronization with other parts of the network was most impaired. Interestingly, these two lobes also contain the two nodes that the consistency-based parenclitic networks revealed to be most affected by MCI. Comparing both results we observe that, despite being in the different lobes, nodes 32 and 61 are not the hub nodes of the functional network, nor are they the nodes whose local properties inside the functional networks were most modified by the disease. This fact indicates that, with the projection of brain dynamics into parenclitic networks, we are evaluating a different effect of the disease on the functioning of the brain network.

\section{Discussion}\label{subsec3}
It is worth mentioning that although we assumed a linear correlation between the channel consistencies, we have not proved that this fitting is the one capturing the real interplay between the consistency of brain regions. Further studies should be devoted to investigating the existence, or co-existence, of higher order correlation functions, although we obtained similar results with a second-order polynomial adjustment (not shown here). In any case, we demonstrated that assuming a linear correlation leads to diﬀerences between groups and allows identifying the nodes whose topological properties are aﬀected the most by the emergence of the disease.

Graph theoretical measures have proven to represent good indicator of the emergence and evolution of a series of brain diseases, an aspect that renders them of enormous practical application. The emergence of brain dysfunction can be quantified using network metrics, which are altered in a disease-specific way \citep{stam2012}. For example, during epileptic seizures, functional brain networks become more regular, modifying their degree distribution and losing part of their modular structure \citep{ponten2007,kaiser2010}. On the contrary, functional networks of schizophrenic patients become more random, with a consequent decrease of both the normalized clustering coefficient and shortest path \citep{micheloyannis2006}. Another example is Alzheimer's disease (AD) which consists of a disconnection syndrome leading to an increased shortest path and decreased network clustering, both leading to a severe impairment of the desirable properties afforded by small-world networks \citep{stam2009}.
The effect of MCI on the structure of functional brain networks has also been
investigated  showing an increased network synchronization and a propensity to enhance long-range connections  \citep{buldu2011,navas2013}. 
Nevertheless, the studies
focused on the coordinated activity between cortical regions and on how the disease alters the topology of
connections within the functional networks. In the current work, we are concerned with another type of consequences of the
emergence of MCI: the loss of a consistent response \citep{uchida2004}, i.e., in our case, the impairment of the ability of a cortical region to behave 
in the same way when undergoing the same cognitive task, despite having different initial conditions. 
We have taken advantage of a new kind of network representation,
the parenclitic network \citep{zanin2013}, where a link between two nodes quantifies the deviations of a certain feature of these nodes
from an expected (healthy) behaviour. We have measured the consistency of $147$ cortical brain regions by means
of magnetoencephalography (see Materials and Methods) and constructed a parenclitic network capturing dysfunctions of
the expected consistency.

The analysis of the topological features of a control group of healthy individuals and a group of 
patients suffering from MCI shows that the parenclitic networks can provide useful information to evaluate how disease alters the consistency between cortical regions. First, we report a higher network strength in the MCI group
when the same number of links are considered in both groups. This fact indicates higher deviations from the expected consistency
performance in the MCI group. Furthermore, we observe the appearance of strong hubs in the patients group, which reveals
that the disease is specially severe at certain cortical regions. Specifically, nodes 32 (frontal lobe) and node 61 (occipital lobe)
are detected to be the focal points of the consistency impairment. Nevertheless, the loss of consistency is not restricted to certain speciﬁc regions. This is reflected by the fact that global network parameters such as the average path length or global eficiency also capture differences between the control and the MCI group. On the contrary, the number of steps needed to go from one node to any other in the parenclitic network is much lower in the MCI group, which indicates that the disease alters network consistency in quite a fundamental way. 

At the local scale, the MCI group shows high values of clustering and local efficiency of the parenclitic network indicating that inconsistency does not emerge in isolated regions but in groups of densely interconnected nodes.

Finally, we have also seen that the networks associated to the control group are more random than those of the MCI group, which has been demonstrated to be a common signature of parenclitic networks in preliminary works \citep{zanin2011}, \citep{zanin2013}. It is important to highlight that the control networks are not purely random in the sense of the definition given by Erd\"{o}s-Renyi \citep{newman2010}, but their network properties are closer to random networks when compared with the MCI. Similarly, the MCI networks are closer to star-like networks, despite having more than one central hub and connections between their peripheral nodes. To the best of our knowledge, this is the first result concerning the construction of parenclitic networks to understand brain functioning and specifically the effect of a neurodegenerative disease. We believe that this technique could be extremely useful to evaluate how different brain diseases deteriorate the normal functioning of the brain activity. At the same time, we must note that parenclitic networks are obtained from the correlation between time series of a brain region, which in turn rely on the physiological properties of the brain. Unfortunately, parenclitic networks do not give any information about the function of brain regions. In that sense, how parenclitic and, more generally, functional networks are related to brain function and to the physiological properties of the brain is still largely unknown.

\section{acknowledgements}
This work was supported by the Spanish Ministry of S\&T [FIS2009-07072], the Community of Madrid 
under the R\& D Program of activities MODELICO-CM [S2009ESP-1691], the Spanish Ministry of economy and competitiveness [PSI2012-38375-C03-01, MTM2012-39101], Fundaci\'on Carolina Doctoral Scholarship Program and Colciencias Doctoral Program 568, as well to a grant from the CAM [S2010/BMD-2460].





\bibliographystyle{elsarticle-num}
\bibliography{CSF_BrainTopoMCILibrary}



\end{document}